\documentclass[10pt,twocolumn,preprintnumbers,amsmath,amssymb,nofootinbib,superscriptaddress,prd]{revtex4-2}

\usepackage[dvipsnames]{xcolor}
\usepackage{bm}
\usepackage{graphicx}
\usepackage[colorlinks,allcolors=RoyalBlue]{hyperref}
\usepackage[normalem]{ulem}

\newcommand{\Schro}{Schr\"{o}dinger}

\newcommand{\bx}{\boldsymbol{x}}
\newcommand{\bk}{\boldsymbol{k}}
\newcommand{\brho}{\bar\rho}

\newcommand{\kp}{k_*}
\newcommand{\ks}{k_*}
\newcommand{\dB}{{\rm dB}}

\newcommand{\GN}{{\rm N}}

\newcommand{\sigshell}{\sigma_*}
\newcommand{\cc}{{\rm c.c.}}

\newcommand{\nD}{3}

\newcommand\wh[1]{{#1}}

\begin{document}
\preprint{FERMILAB-PUB-25-0204-T}
\title{Interference with Gravitational Instability: Hot and Fuzzy Dark Matter}

\author{Rayne Liu}

\affiliation{Kavli Institute for Cosmological Physics, Enrico Fermi Institute, and Department of Astronomy \& Astrophysics, University of Chicago, Chicago IL 60637
}

\author{Wayne Hu}
\affiliation{Kavli Institute for Cosmological Physics, Enrico Fermi Institute, and Department of Astronomy \& Astrophysics, University of Chicago, Chicago IL 60637
}

\author{Huangyu Xiao}
\affiliation{Kavli Institute for Cosmological Physics, Enrico Fermi Institute, and Department of Astronomy \& Astrophysics, University of Chicago, Chicago IL 60637
}
\affiliation{Theory Division, Fermi National Accelerator Laboratory, Batavia, IL 60510, USA}
\begin{abstract}
Wave or fuzzy dark matter produced with high momenta behaves in many ways like hot particle dark matter while also possessing seemingly different phenomenology due to wave interference. We develop wave perturbation theory to show that white noise density fluctuations generated by the interference of high-momenta waves are gravitationally unstable in the usual way during matter domination above the free streaming scale and stabilize below the free streaming scale, much like the analogous effects for massive neutrinos in hot dark matter.   We verify and illustrate these effects in the density power spectra of  Newtonian \Schro-Poisson simulations.  In the cosmological context, this would cause a gradual suppression of the initial white noise isocurvature perturbations below the free streaming scale at matter radiation equality, unlike cold dark matter isocurvature fluctuations, and virial stability of dark matter halos.
\end{abstract}

\date{\today}

\maketitle

\section{Introduction}

Wave dark matter candidates span the ultralight regime of masses $m\lesssim 30$\,eV where occupation numbers are high and the field behaves as a classical wave with macroscopic de Broglie wavelengths (see \cite{Hui:2021tkt} for a review).   The prototypical candidate is the QCD axion  \cite{Peccei:1977hh,PhysRevLett.40.223,PhysRevLett.40.279,PhysRevLett.43.103, Peccei:2006as}
which can be produced after inflation due to the variations in the misalignment angle across horizon scale patches
\cite{Abbott:1982af, Dine:1982ah,Preskill:1982cy}. While the QCD axion born in this way is mostly cold, with initial wavelengths of order the Hubble scale at birth, more generally the causal production of such ultralight bosons could generate a quasirelativistic or even ultrarelativistic momentum distribution \cite{Vaquero:2018tib,OHare:2021zrq,Benabou:2024msj}. Such cases are the wave analogue of warm and hot particle dark matter, which we refer to as warm or hot fuzzy dark matter (FDM).  Here fuzzy refers to the mass range where the de Broglie wavelength is on astrophysical scales \cite{Hu:2000ke}.
Furthermore, even an initially cold distribution can acquire a high momentum component due to gravitational or self-interactions \cite{Gough:2022pof,Gorghetto:2024vnp}.  

As with warm and hot particle dark matter, the distribution of wave momenta affects the formation of small scale structure.   On small scales, the high momenta wave components free stream and erase the initial density perturbations from inflation \cite{Amin:2022nlh} as well as the isocurvature number density fluctuations from their coherent generation in causal patches \cite{Liu:2024pjg}.  However, free streaming waves also leave in their wake transient interference effects in the energy density, which themselves are white noise distributed on large scales \cite{Amin:2022nlh,Liu:2024pjg,Ling:2024qfv}.   For the warm FDM case we showed that the impact on the adiabatic modes differs from cold FDM by a logarithmic correction and bounds on isocurvature modes are largely unaffected \cite{Irsic:2019iff}.  Moreover, the wave and particle pictures for these effects, while seemingly different, predict the same phenomenology. 

On the other hand, in the hot FDM case, the initial momenta can in principle be so large that the free streaming scale parametrically differs from the Jeans scale of the cold FDM case \cite{Amin:2022nlh}.  Here the interference effects that remain on small scales after free streaming has eliminated adiabatic fluctuations and may play a more important role in the gravitational collapse of density fluctuations. Nevertheless, such interference effects are the wave analogue of a coarse grained velocity dispersion of particle dark matter, which in that case impedes rather than seeds gravitational collapse, and instead results in virial equilibrium below the effective Jeans scale.  In this sense, interference effects correspond to the fine grained phase space density fluctuations of collisionless particle dark matter (see e.g.~\cite{Gough:2022pof}) and should not be equated with the number density fluctuations of cold dark matter isocurvature modes \cite{Liu:2024pjg}.

In this work we study the gravitational stability of wave-interference generated density fluctuations in this hot FDM regime and the role of interference in impeding gravitational collapse.  We find that these effects halt the gravitational growth of perturbations under the free streaming scale just as in particle hot dark matter. The differences arise mainly from the evolution of the free streaming scale itself.  In the hot FDM case, this would lead to a gradual relative suppression of isocurvature fluctuations on small scales as well as stability in virialized objects for initially cold FDM.

This work is organized as follows. In Sec.~\ref{sec:hot_fuzzy} we discuss the equations and timescales governing hot and fuzzy dark matter in the non-relativistic regime. 
 In Sec.~\ref{sec:perturbation_theory}, we
present the perturbative formalism for evolving the interference density fluctuations of high momentum waves and show that their growth ceases below the free streaming scale whereas larger scales grow on the dynamical time. 
In Sec.~\ref{sec:interference}, we verify this behavior with numerical simulations for an illustrative case that possesses closed form perturbative solutions as well as white noise field initial conditions. In Sec.~\ref{sec:discussion}, we discuss the cosmological implications of these findings for hot and fuzzy dark matter.

\section{Hot and Fuzzy Dark Matter}\label{sec:hot_fuzzy}

We consider the interplay between gravitational interactions and wave interference of a free scalar field $\phi$.     In the non-relativistic regime, the field takes the form 
\begin{equation}
     \phi = \frac{1}{\sqrt{2m}}\left(\psi e^{-imt}+\cc     \right),
\end{equation}
where the wavefunction $\psi$  evolves under the \Schro-Poisson  equation \cite{Hu:2000ke}
\begin{align}\label{eq:schro-poisson}
    i(\partial_t + \frac{3}{2} H){\psi} &= -\frac{1}{2m}\nabla^2\psi+m\Psi\psi, \\
    \nabla^2 \Psi &= 4\pi G (\rho-\brho). \nonumber
\end{align}
Here $m$ is the scalar mass, $H=d\ln a/dt$ is the expansion rate, $\rho$ is the energy density, $\brho$ is its spatial average, and $\Psi$ is the gravitational potential. If we consider $\psi$ as the dominant form of matter
\begin{equation}
\label{eq:rho}
    \rho({\bx},t) \equiv m |\psi({\bx},t)|^2.
\end{equation}

The Fourier transform of $\psi$ at some initial reference time denoted as $t=0$
\begin{eqnarray}
\psi_i(\bk)=\psi(\bk,t=0) &=& \int d^3 x e^{-i \bk \cdot\bx  }\psi(\bx,0)
\end{eqnarray}
determines the momentum distribution through its power spectrum
\begin{equation}
\langle \psi_i^*({\bk}) \psi_i({\bk}')\rangle = (2\pi)^3 \delta({\bk}-{\bk}') P_{i}(k).
\label{eq:Pi}
\end{equation}
For the density modes
\begin{eqnarray}
\rho(\bk,t) &=&  m \int  \frac{d^3 k'}{(2\pi)^3} \psi^*(\bk'-\bk,t) \psi(\bk',t),
\label{eq:densitymode}
\end{eqnarray}
and the initial density power spectrum is defined by
\begin{eqnarray}
\langle \rho^*(\bk,0)\rho(\bk',0)\rangle
&=& (2\pi)^3 \delta({\bk}-{\bk}') P_{\rho}(k,0), \\
P_{\rho}(k,0)  &=&
m^2 \int \frac{d^3 k_a}{(2\pi)^3} 
P_i({\bk}_a-{\bk})P_i({\bk}_a).\nonumber
\end{eqnarray}
Notice that even if the field has a vanishingly small mean or equivalently its power spectrum $P_i(0)=0$, the density field has a finite mean and power $P_\rho(0,0)\ne 0$.  For field power spectra where $k^3 P_i$ peaks at high $k =\kp$, which we refer to as a hot distribution, the density power spectrum becomes white noise $P_\rho(k,0)=P_\rho(0,0)$  for $k\ll \kp$.   These low $k$ density fluctuations therefore are purely the consequence of the interference of high $k$ field modes.

Note $\brho$ need not represent the cosmological mean density nor does the hot distribution need to be so at dark matter production.   For example, in collapsed objects such as dark matter halos with higher local density, for sufficiently large mass, the momentum distribution is also technically hot due to the velocity dispersion of the dark matter.   Here the wavelengths of the dominant waves are much smaller than the size of the system.

Nonetheless, wave dark matter can also be produced with a hot momentum distribution 
in the early universe. 
To be a substantial fraction of the dark matter, its production must have occurred before matter-radiation equality so that it is non-relativistic in the matter dominated epoch.
For example, for axionlike dark matter, 
the momentum distribution of the field peaks at or below the horizon scale at formation and the field lacks correlations on larger scales.
While the free streaming of the dark matter during radiation domination alters the shape of the momentum spectrum, it remains sufficiently peaked near the redshifted original momentum for the density  fluctuation power spectrum from interference of these modes to remain white \cite{Amin:2022nlh,Liu:2024pjg,Ling:2024qfv,Long:2024imw}.
Free streaming of the waves during radiation domination also eliminates adiabatic density fluctuations, leaving this white noise interference spectrum in their wake.

Given a characteristic $\kp$, we can immediately identify three time scales in 
Eq.~(\ref{eq:schro-poisson}): the Hubble timescale $H^{-1}$, the free oscillation timescale $t_\dB = m/\kp^2$, which is the wave crossing time of the de Broglie wavelength of $\kp$, and the  dynamical timescale $t_\GN=1/\sqrt{G \brho}$.  For cold dark matter, small scale density fluctuations that are initially adiabatic or isocurvature  would grow on the $t_\GN$ timescale in the matter dominated epoch.  
Next we turn to the question of whether white noise fluctuations from wave interference in the hot FDM regime where $t_\GN \gg t_\dB$ are gravitationally unstable on the same timescale.

\section{Wave Perturbation Theory}\label{sec:perturbation_theory}

Extending the techniques  of Ref.~\cite{Capanelli:2025nrj}, we consider the initial growth of white noise isocurvature density fluctuations from  interference of high momentum  waves in perturbation theory.  To highlight the role of interference and illustrate the physical properties in a simple setting, we ignore the expansion and take $H t_\GN \ll 1$.   Cosmologically this situation can occur in local regions where the average density is larger than the cosmic mean, e.g.~a collapsed dark matter halo.  We return to discuss the similarities and differences with the case of small fluctuations around the cosmic mean in Sec.~\ref{sec:discussion}.

For the typical wave momentum scale $\kp$, we take the case where $t_\GN \gg t_\dB$, i.e.~the wavelength is much smaller than the FDM Jeans scale \cite{Khlopov:1985fch,Hu:2000ke}
\begin{equation}
\epsilon_G \equiv \frac{16 \pi G\brho m^2}{\kp^4} = 16\pi \left( \frac{t_\dB}{t_\GN} \right)^2 \ll 1,
\end{equation}
so that density modes around $\kp$ are gravitationally stable.   We call this case a hot FDM distribution, in contrast with $\epsilon_G={\cal O}(1)$ for a warm, and $\epsilon_G\gg 1$ for a cold, FDM distribution.

In this case, for the wavemodes near $\kp$, the \Schro-Poisson equation in Fourier space describes the free propagation of the initial waves
\begin{equation}
\psi_0(\bk,t) = \psi_i(\bk) e^{-i\frac{k^2}{2m} t}.
\label{eq:psi0}
\end{equation}
Note that free evolution just changes the phase of the wave and does not change their momentum distribution.  For $\epsilon_G\ll 1$, we can therefore treat the effect of gravity on the wavefunction as a perturbation around $\psi_0$.

As with the density perturbation, the self-gravitational potential fluctuation is a convolution of field momenta modes 
\begin{equation}
\Psi(\bk,t) = -\frac{4\pi G m}{k^2}
\int \frac{d^3 k'}{(2\pi)^3} \psi^*(\bk'-\bk,t)\psi(\bk',t) ,
\end{equation} 
and we can iteratively solve for its effect by taking successive approximations for $\psi$  to evaluate this gravitational source.
More specifically, with 
\begin{equation}
S(\bx,t)=m \Psi(\bx,t)\psi(\bx,t)
\end{equation}
as the  source, the field perturbation $\psi = \psi_0+\delta\psi$ evolves according to Eq.~(\ref{eq:psi0}) as
\begin{eqnarray}
    i\dot{\delta\psi}&&= -\frac{1}{2m}\nabla^2\delta\psi+S(\bx ,t).
    \label{eq:psiiteration}
\end{eqnarray}
We use the $n$th order solution for $\psi$ to evaluate $S$ and find the ($n+1$)th order correction.  In particular the first order correction is 
\begin{eqnarray}
\psi_1(\bk,t) &=&  i(4\pi G m^2)  e^{-i\frac{k^2}{2m} t}\left[ \int_0^t dt' e^{i \frac{k^2}{2m} t' }   \right. \nonumber\\
&&\times \left.
 \int\frac{d^\nD k_1}{(2\pi)^\nD}
 \int \frac{d^\nD k_2}{(2\pi)^\nD}
e^{i \frac{k_3^2 - k_1^2 - k_2^2}{2m} t'} \right. \nonumber\\
&&\times  \left.\frac{1}{|\bk-\bk_1|^2} \psi_i(\bk_1) \psi_i(\bk_2) \psi_i^*(\bk_3) \right],
\end{eqnarray}
where $\bk_3 = \bk_1+\bk_2-\bk$. 
Furthermore, the unperturbed density field $\rho_0$ is given by Eq.~(\ref{eq:densitymode}) using $\psi_0$ and the first order correction by
\begin{eqnarray}
\rho_1(\bk,t) &=& m  \int\frac{d^3 k'}{(2\pi)^3} \left[ \psi_1^*(\bk',t)\psi_0(\bk'+\bk,t) \right. \nonumber\\
&& \left. + \psi_0^*(\bk'-\bk,t) \psi_1(\bk',t) \right].
\end{eqnarray}
The perturbation to density power spectrum 
\begin{equation}
\delta P_{\rho}(k,t) \equiv P_{\rho}(k,t)-  P_{\rho}(k,0)
\end{equation}
receives its leading order contribution from 
\begin{equation}
 \langle \rho_0^*(\bk',t) \rho_1(\bk,t) \rangle +\cc
 \approx
 (2\pi)^3 \delta(\bk-\bk')\delta P_\rho(k,t),
 \end{equation}
where
\begin{eqnarray}
\delta P_\rho(k,t) &\approx& 16\pi G m^4 \! \int\frac{d^\nD k_a}{(2\pi)^\nD}
\!\! \int \frac{d^\nD k_b}{(2\pi)^\nD}  
 P_i(k_a) P_i(k_b)  P_i(k_c) \nonumber\\
&&\times\left(\frac{1}{k^2} + \frac{1}{|\bk_b +\bk_c|^2 } \right) \frac{4 m}{\Delta q^2} \sin^2(\frac{\Delta q^2}{4 m} t)
\label{eq:deltaPrho}
\end{eqnarray}
with $\bk_c=\bk+\bk_a$ and 
\begin{eqnarray}
\Delta q^2 & =&  k_c^2 + (\bk_b+\bk)^2 - k_a^2 - k_b^2 \nonumber\\
&=& 2(\bk_a\cdot \bk + \bk_b\cdot\bk + k^2) = 2(\bk_b+\bk_c)\cdot \bk.
\end{eqnarray}
Notice that $\Delta q^2$ carries the phases between the perturbed $\psi_1(\bk+\bk_b)$ mode and the three unperturbed $\psi_0$ modes $k_{a,b,c} \sim \kp$ that it is correlated with.   
If $\Delta q^2 t/4 m\ll 1$ these phases are coherent and the density perturbations from wave interference will grow as $\delta P_\rho \propto t^2$.

The three power spectra in Eq.~(\ref{eq:deltaPrho}) reflect the 6 point nature of the leading order $\delta P_\rho$ term and the assumption that connected correlators of $\psi_i$ vanish beyond the 2 point.  Despite the seemingly large number of possible paired contractions between modes, with permutation symmetry, conjugation and statistical isotropy, only two types of terms survive, distinguished by the wavenumber of the gravitational potential $\Psi$:
$\bk$ or $\bk_b+\bk_c$.  Notice that for the former, this is the usual case where the density fluctuation and $\Psi$ share the same wavenumber in the Jeans instability analysis. In the latter case, which we shall see is subdominant, it is mainly the short wavelength $k_*$ gravitational potential fluctuations combined with short wavelength $\psi_0$ modes that impact the long wavelength $k$ density mode in a squeezed configuration.

In fact the timescale for the dominant growth is simply the dynamical time, as one might expect from the association above, and this is true for any shape of  the field spectrum $k^3 P_i$ as long as it is dominated by high momenta.   To show this, let us re-express Eq.~(\ref{eq:deltaPrho}) using the dynamical time as
\begin{eqnarray}
\lim_{t\ll m/\Delta q^2} \frac{\delta P_\rho(k,t)}{P_\rho(k,0)} &= & 
4\pi A \left(\frac{t}{t_\GN} \right)^2,
\end{eqnarray}
where 
\begin{eqnarray} 
A &=&   \frac{4 m^3}{P_\rho(k,0)\brho}  \int\frac{d^\nD k_a}{(2\pi)^\nD}
 \int \frac{d^\nD k_b}{(2\pi)^\nD} P_i(k_a) P_i(k_b) P_i(k_c)  \nonumber\\ 
 &&
\times \frac{\Delta q^2} {4}  \left(\frac{1}{k^2} +\frac{1}{ |\bk_b +\bk_c|^2}\right).
\label{eq:Afull}
 \end{eqnarray}
For $k\ll \kp$, we can evaluate the dominant $1/k^2$ term directly.  Notice that the integral of the $\bk_b\cdot\bk$ piece of $\Delta q^2$ is zero  by symmetry  and the remaining integral over $\bk_b$ is the same as that for $\brho$ so
\begin{eqnarray}
A\approx 
 \frac{2  m^2}{P_\rho(k,0)}  \int\frac{d^\nD k_a}{(2\pi)^\nD}
  P_i(k_a) P_i(k_c) 
 \frac{ \bk_a\cdot \bk +k^2 }{k^2}.
 \end{eqnarray}
Next since $k_c = \sqrt{ k_a^2 + 2 \bk_a\cdot \bk + k^2}$, we can Taylor expand
\begin{equation}
P_i(k_c) \approx P_i(k_a) + \frac{d P_i(k_a)}{d k_a}
\frac{\bk_a \cdot \bk}{k_a}+ \ldots
\end{equation}
and again integrate over angles
\begin{equation}
A\approx 
 \frac{2  m^2}{P_\rho(k,0)}  \int dk_a \frac{k_a^2P_i^2(k_a) }{2\pi^2}
 \left( 1 + \frac{1}{3} 
  \frac{d\ln P_i(k_a)}{d\ln k_a}
 \right).
\end{equation}
We can then integrate by parts assuming zero boundaries at $k=0,\infty$
 \begin{eqnarray}
\frac{1}{3}{\int dk_a k_a^{3}  P_i(k_a) \frac{ d P_i}{d k_a}}&=&
-\frac{1}{2}\int dk_a k_a^{2}  P_i^2(k_a) 
\nonumber\\
&\approx &-\frac{\pi^2}{m^2}
P_\rho(0,0).
 \end{eqnarray}
Finally since $k\ll \kp$, $P_\rho(k,0) \approx P_\rho(0,0)$ is constant and white.
Putting this together, we have
\begin{equation}
A \approx 1
\end{equation}
independently of the specific shape of $P_i(k)$.

Notice that in this case $\delta P_{\rho}/P_\rho$ reaches ${\cal O}(1)$ on the dynamical time scale $t_\GN$
\begin{equation}
\lim_{t\ll m/k k_*} \frac{\delta P_\rho}{P_\rho} = 4\pi (t/t_\GN)^2
\end{equation}
as one would expect from cold dark matter.  Even though the density fluctuations at these long wavelengths are still originally generated by the interference of high momenta waves, they seed gravitational instability of overdense regions in the usual way of causing gravitational infall and clustering of the dark matter.

On the other hand, once $\Delta q^2 t/4 m \gtrsim1$ in Eq.~(\ref{eq:deltaPrho}), the contributions to the density power from such $k$ configurations oscillate rather than grow.   Given that $k_b \sim k_c \sim k_*$, $\Delta q^2 \sim k k_*$ and this growth will saturate when
\begin{equation}
t \sim \frac{ m}{k \kp}
\end{equation}
due to phase decoherence. 
Notice that this occurs when the free streaming length of the $\kp= mv_*$ waves overtakes the perturbation wavelength 
\begin{equation}
\lambda_{\rm fs} = \frac{\kp}{m} t 
\sim k^{-1}.
\end{equation}
The exception is for special configurations where $\Delta q^2 \rightarrow 0$ where $\bk_a \approx \bk_c \approx -\bk_b$ but these configurations are a small fraction of the phase space. Likewise, except for these special configurations the $1/k^2$ term in Eq.~(\ref{eq:deltaPrho}) will dominate for $k\ll \kp$ (see Eq.~(\ref{eq:Ashell}) for an explicit example).

After $t\gg k k_*/m$ or when the density perturbation is well under the free streaming length, the power spectrum growth saturates at an amplitude
\begin{equation}
\lim_{t\gg m/k k_*} \frac{\delta P_\rho}{P_\rho} \sim 16 \pi 
\left( \frac{m}{k \kp t_\GN} \right)^2 = \epsilon_G \left( \frac{\kp}{k} \right)^2.
\end{equation}
This happens because the decoherence over time of the phases of the high momentum waves that compose $\Delta q^2$  prevents a coherent gravitational response to the initial density perturbation.
We next turn to specific illustrations of these generic expectations.

\begin{figure}[t]
    \centering
\includegraphics[width=\linewidth]{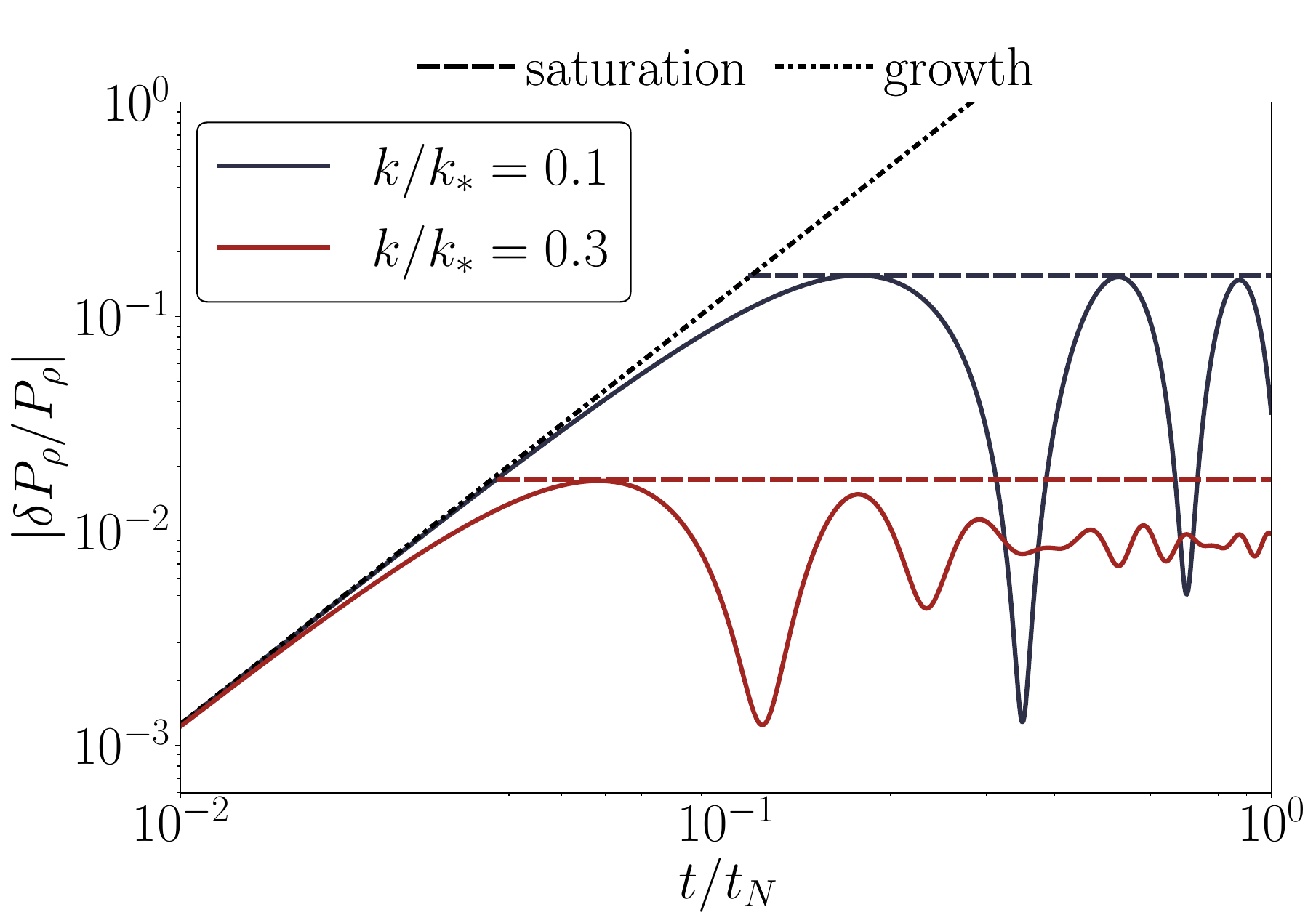}
    \caption{Perturbation theory predictions for gravitational growth and interference saturation of the initial density power spectrum $P_\delta(k,0)$ for $k$ modes that cross the free streaming scale at different times.  High $k$ modes cross earlier and their growth saturates at a low fractional change and oscillate thereafter.  Scaling predictions in the growth and saturation regime from Eq.~(\ref{eq:scalings}) are shown in dashed lines.  Analytic predictions from Eq.~(\ref{eq:dPtheory}) here are for an infinitesimal width shell of field momentum modes at $k=\kp$ and $t_\dB/t_\GN \approx 0.006$.
        }    
    \label{fig:time}
\end{figure}

\section{Interference impeded growth}\label{sec:interference}

In this section, we provide a fully worked example of how interference impedes the gravitational growth of density fluctuations which are themselves generated from the interference of high momenta waves.   In Sec.~\ref{sec:shell}, we characterize this momentum distribution with a shell configuration where perturbation theory solutions can be expressed in closed form.  In Sec.~\ref{sec:sims} we test these predictions in the perturbative regime and show that the characteristic scale between growth and saturation is the free streaming scale.

\subsection{High momentum shell}
\label{sec:shell}

Following \cite{Capanelli:2025nrj}, we take an initial shell configuration where the wave modes of the field are exclusively at high momentum $\epsilon_G\ll 1$ around a characteristic scale $\kp$
\begin{equation}\label{eq:initial-condition}
   \psi_i(\bk) \propto 
e^{ - \frac{(k-\ks)^2}{2\sigshell^2} + i \alpha_{\boldsymbol{k}}},
\end{equation}
with the Gaussian shell width $\sigshell \ll \ks$ and uniform random phases $\alpha_{\bk}$.  We normalize the modes so that 
\begin{equation}
\psi_{\rm rms}^2 = \int \frac{dk}{k} \frac{k^3 P_i}{2\pi^2}=\frac{\brho}{m}.
\end{equation}
For pedagogy, we start with this example rather than a power law $P_i$ since $\sigshell$ provides a way of strictly eliminating low momentum waves without a large dynamic range of wavenumbers for the simulations of the next section. Low wavenumber density modes then also strictly come from the interference of high momentum waves as long as $\sigshell \ll \kp$.  Nonetheless for $k\ll \sigshell$, these density modes are white-noise distributed just like in the power law  $P_i$  model.
We return to the white noise $P_i$ model in Sec.~\ref{sec:sims}.

The limiting case of $\sigshell\rightarrow 0$ is also interesting in that perturbation theory predictions can be expressed in closed form.  Here the initial field and density power spectra take the form 
\begin{eqnarray}
\lim_{\sigshell\rightarrow 0} P_i(k) & = &\delta(k-\kp) \frac{2 \pi^2 }{k^2} \frac{\bar \rho}{m}, \nonumber\\
\lim_{\sigshell\rightarrow 0} P_{\rho}(k,0) &=& 
\begin{cases}
 \frac{\pi^2}{k \kp^2}\bar\rho^2, & (k\le 2 \kp)\\
 0, & (k> 2 \kp)
 \end{cases}.
 \label{eq:initialP}
\end{eqnarray}
While for a finite $\sigma_*$, the density power spectrum would turn over from $P_\rho \propto k^{-1}$ to become white for $k \ll \sigma_*$, in this limiting case it does not, but results in terms of $\delta P_\rho/P_\rho$ are insensitive to this difference as we shall see in the next section where we compare this idealization to finite width simulations.

In this limit, we can explicitly derive the predictions in the early time growth regime of Eq.~(\ref{eq:Afull}) and show that above the free streaming scale density perturbations grow on the dynamical timescale.  With Eq.~(\ref{eq:initialP}), the delta functions in the three power spectra $P_i(k_{a,b,c})$ allow us to immediately perform the integrals over $k_a$, $k_b$ and the polar angle $\bk_a \cdot \bk = k \kp x_a$ using
$\delta(k_c-k_*)= \delta(x_a+k/2\kp)/k$, which gives
 \begin{equation}
\Delta q^2 = 2 k \kp ( x_a + x_b + k/\kp)
=  k^2  + 2k\kp x_b,
 \end{equation}
 where $\bk_b \cdot \bk = k \kp x_b$.  The  integrals  over the azimuthal angles and $x_b$ then become
\begin{eqnarray} 
A &=&    \int_0^{2\pi} \frac{d\phi_a}{2\pi}
\int_0^{2\pi} \frac{d\phi_b}{2\pi} \int_{-1}^{1} dx_b
\left(\frac{1}{2}+ \frac{\kp}{k} x_b \right)\nonumber\\
&&\times
 \left(1 +\frac{k^2}{ |\bk_b +\bk_c|^2}\right).
\label{eq:Adelta}
\end{eqnarray}
The leading order  term in the second line obviously integrates to unity as expected.   The subdominant term can be simplified by a change of angular basis to orient the polar angle of $\bk_b$ as $\bk_b \cdot \bk_c = \kp^2 \cos\tilde\theta_b$ and its azimuthal angle $\phi_b$  measured relative to that of $\bk$
\begin{eqnarray}
|\bk_b + \bk_c|^2 &=& 2\kp^2(1+\cos\tilde \theta_b),\nonumber\\
\frac{1}{2}+ \frac{\kp}{k} x_b &=& \frac{1}{2}(1+ \cos\tilde\theta_b)+\frac{\kp}{k}\sin\tilde\theta_b\cos\tilde\phi_b,
\end{eqnarray}
where we have used that $\bk\cdot\bk_c=
-\bk\cdot\bk_a$ given that these modes form an isosceles triangle for $\sigshell\rightarrow 0$.
The net result is 
\begin{equation}
A = 1+ \frac{1}{2}\frac{k^2}{\kp^2}.
\label{eq:Ashell}
\end{equation}
Since we consider $k\ll \kp$ we can ignore the subdominant term as expected.  With $A=1$ all such scales initially  grow on the same $t_\GN$ timescale.  

Once the free streaming scale crosses the density perturbation wavelength, gravitational growth ceases.
It is also possible to integrate the dominant term past the growth epoch into this saturation epoch:
\begin{eqnarray}
\lim_{\sigma_*\rightarrow 0}
\frac{\delta P_\rho}{P_\rho } &\equiv & {4\pi } \frac{t_\dB^2}{t_\GN^2}
F\left(\frac{k}{\kp},\frac{t}{t_\dB}\right),
\label{eq:dPtheory}
\end{eqnarray}
where
\begin{eqnarray}
F(\kappa,\tau) &=& 
 8\int_{-1}^{1} dx_b
\frac{\sin^2\left( 
(\kappa+ 2 x_b) \kappa\tau/4 \right)}{\kappa^4 + 2 x_b \kappa^3}
 \nonumber\\
&=& \frac{2}{\kappa^3}
\left(
2 \tanh^{-1}[\kappa/2]+ {\rm Ci}[\tau(2-\kappa) \kappa/2] \right.
\nonumber\\
&&\left. - {\rm Ci}[\tau(2+\kappa) \kappa/2]
\right)
\end{eqnarray}
and Ci is the cosine integral.
Note that
$F(\kappa,\tau\rightarrow 0)= \tau^2 = (t/t_\dB)^2 $ or equivalently $A=  (t_\dB/t)^2F = 1$ as expected. 
Conversely $F$ reaches peak growth at
$\kappa \tau = k \kp t/m =\pi$ at an amplitude 
of $F=4 \kappa^{-2}$ and  oscillates thereafter.  For a finite $\sigshell$ eventually the various contributing configurations of $\Delta q^2$ within the width $\sigshell$ will oscillate out of phase and incoherently superimpose with each other.   Therefore, more generally we expect the density perturbations to grow and then saturate with the parametric scaling
\begin{equation}
\label{eq:scalings}
\frac{\delta P_\rho}{P_\rho} \approx
\begin{cases}
4\pi (t/t_N)^2, & k t\le 2m/ \kp \\
16\pi  \left( \frac{m}{k \kp t_N} \right)^2, & k t> 2m/\kp
\end{cases}.
\end{equation}

In Fig.~\ref{fig:time}, we show an example of the time evolution for $\delta P_\rho/P_\rho$ in the $\sigshell\rightarrow 0$ case.  All modes start with $4\pi(t/t_\GN)^2$ growth but the free streaming scale crosses the higher $k$ mode first and growth saturates thereafter.   The further time evolution oscillates but the specific pattern depends on the superposition of the oscillating components in $\Delta q^2$.   We also show that the simple scalings from Eq.~(\ref{eq:scalings}) for the growth and saturation regimes characterize the qualitative behavior well.

\begin{figure}[t]
    \centering
\includegraphics[width=\linewidth]{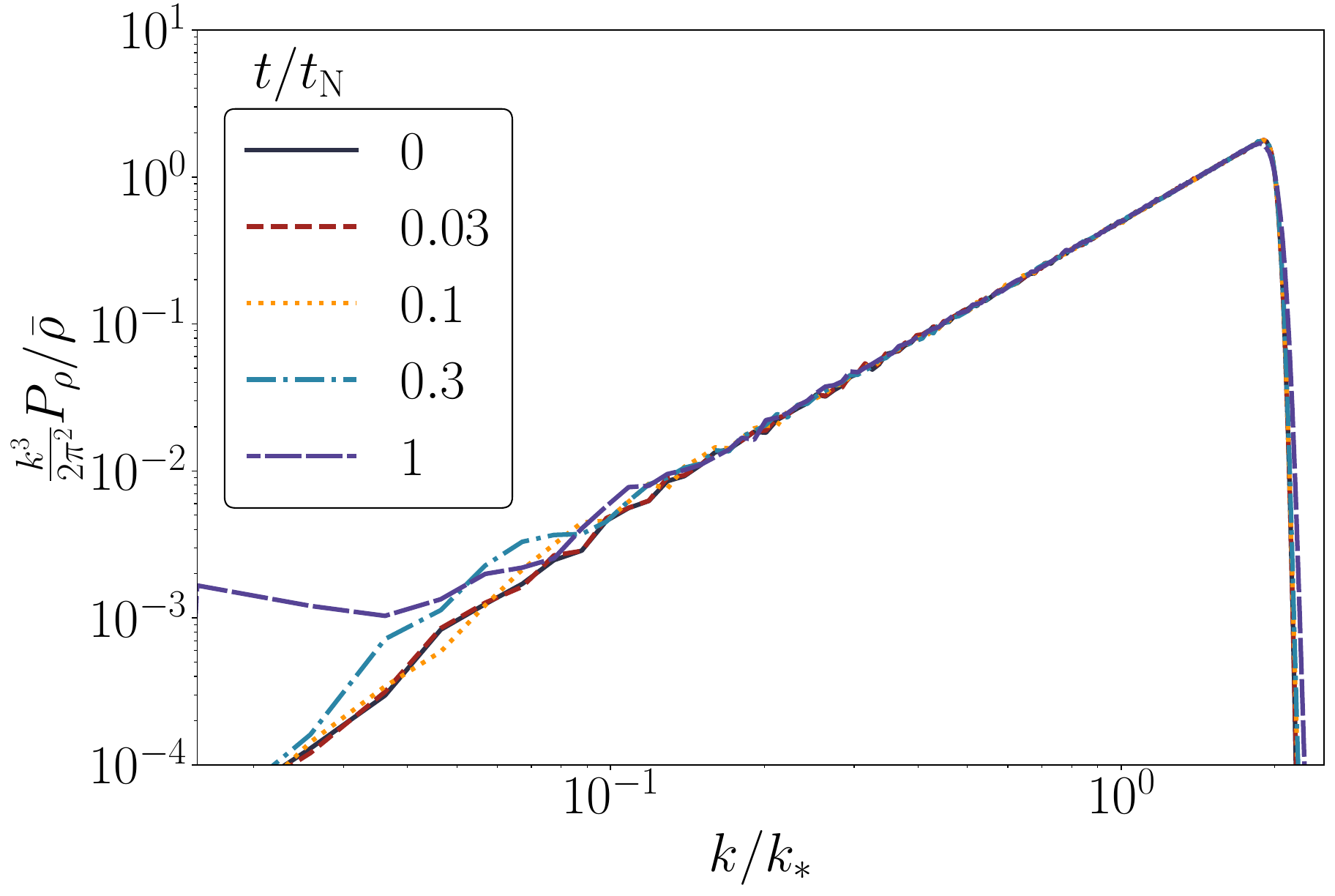}
    \caption{Density power spectrum in simulations for a high momentum shell of field fluctuations with width $\sigshell/\kp=0.05$ and $t_\dB/t_\GN=0.006$. High $k$ modes below the free streaming scale do not grow whereas above the free streaming scale they grow on the dynamical timescale $t_\GN$ as expected.}
    \label{fig:power}
\end{figure}

\begin{figure}[t]
    \centering
\includegraphics[width=\linewidth]{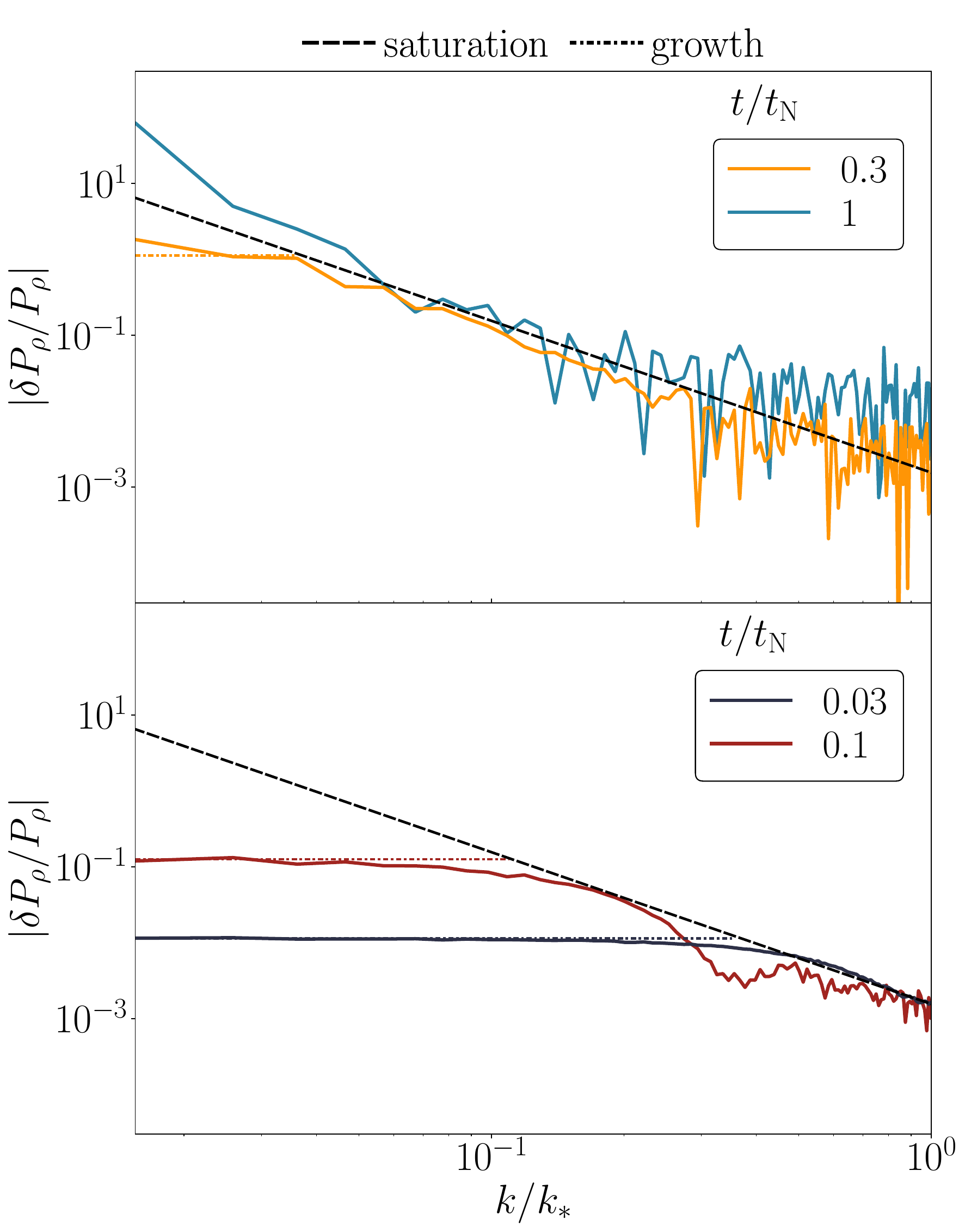}
    \caption{Gravitational growth of the density power spectrum in the shell simulations (solid lines).  In the perturbative regime (bottom, $t/t_\GN \lesssim 0.1$) the growth above, and saturation below, the evolving free streaming are well predicted by the perturbation theory scalings (dot-dashed and dashed lines) whereas the growth after a dynamical time (top, $t/t_\GN \gtrsim 0.3$) is exponential in simulations whereas saturation below the free streaming scale persists.  Simulation parameters are the same as in Fig.\ \ref{fig:power}.
    }
    \label{fig:dP}
\end{figure}

\subsection{Simulations}
\label{sec:sims}

Next we test these predictions against simulations of the \Schro-Poisson system and extend them into the nonlinear regime.   We use the publicly available 
PyUltraLight code \cite{Edwards:2018ccc} which employs the pseudo-spectral approach to evolve the initial conditions forward in a non-expanding background.  The simulation results are independent of the specific parameters such as $m$ or $\brho$ as long as wavenumbers are scaled to $\kp$ and time scales to $t_\GN$ for a given $t_\dB/t_\GN$, which quantifies how hot the initial momenta are.

As an example we take a shell width $\sigshell/\kp=0.05$ and $t_\dB/t_\GN=0.006$ using a $256^3$ simulation grid. 
In Fig.~\ref{fig:power} we show the evolution of $P_\rho$ itself.   Notice that even after a dynamical time $t_\GN$, the high $k$ modes evolve negligibly whereas they would be predicted to evolve by ${\cal O}(1)$ under gravity alone.   As expected from perturbation theory, it is only for the lowest $k$ modes, whose wavelengths are larger than the free streaming scale at $t_\GN$, that the growth is appreciable.   For even later times, the power spectrum on scales below the free-streaming scale at $t_\GN$ remains nearly constant.

To test the perturbation theory predictions for the $\delta P_\rho/P_\rho$ change due to gravity   more directly we define it in the simulations as 
\begin{equation}
\frac{\delta P_\rho}{P_\rho} = \frac{P_\rho - P_{\rho_0}}{P_{\rho_0}},
\end{equation}
where $P_{\rho_0}$ is the power spectrum constructed from the free field $\psi_0$ solution in Eq.~(\ref{eq:psi0}) with the same initial phases as the simulation realization. 
In Fig.~\ref{fig:dP}, we compare these  changes to the analytic scaling relations (\ref{eq:scalings}) from perturbation theory.  For $t/t_\GN \lesssim 0.1$ (bottom panel) perturbation theory fully applies and both the growth and saturation are well predicted by the scalings.
For $t/t_\GN\gtrsim 0.3$ (top panel), the modes above the freestreaming scale grow faster than $t^2$.  This is expected since beyond perturbation theory, gravitational instability in a non-expanding medium is exponential rather than power law.  Notice that even at these late times, perturbations below the free streaming scale still oscillate rather than grow and the perturbation theory scalings still provide a reasonable estimate for the mean level.  

We have checked that these scaling results for growth and saturation in the perturbative regime do not depend  on the arbitrary width $\sigshell$ chosen, though the oscillatory patterns in the saturation regime do.

\begin{figure}[t]
    \centering
\includegraphics[width=\linewidth]{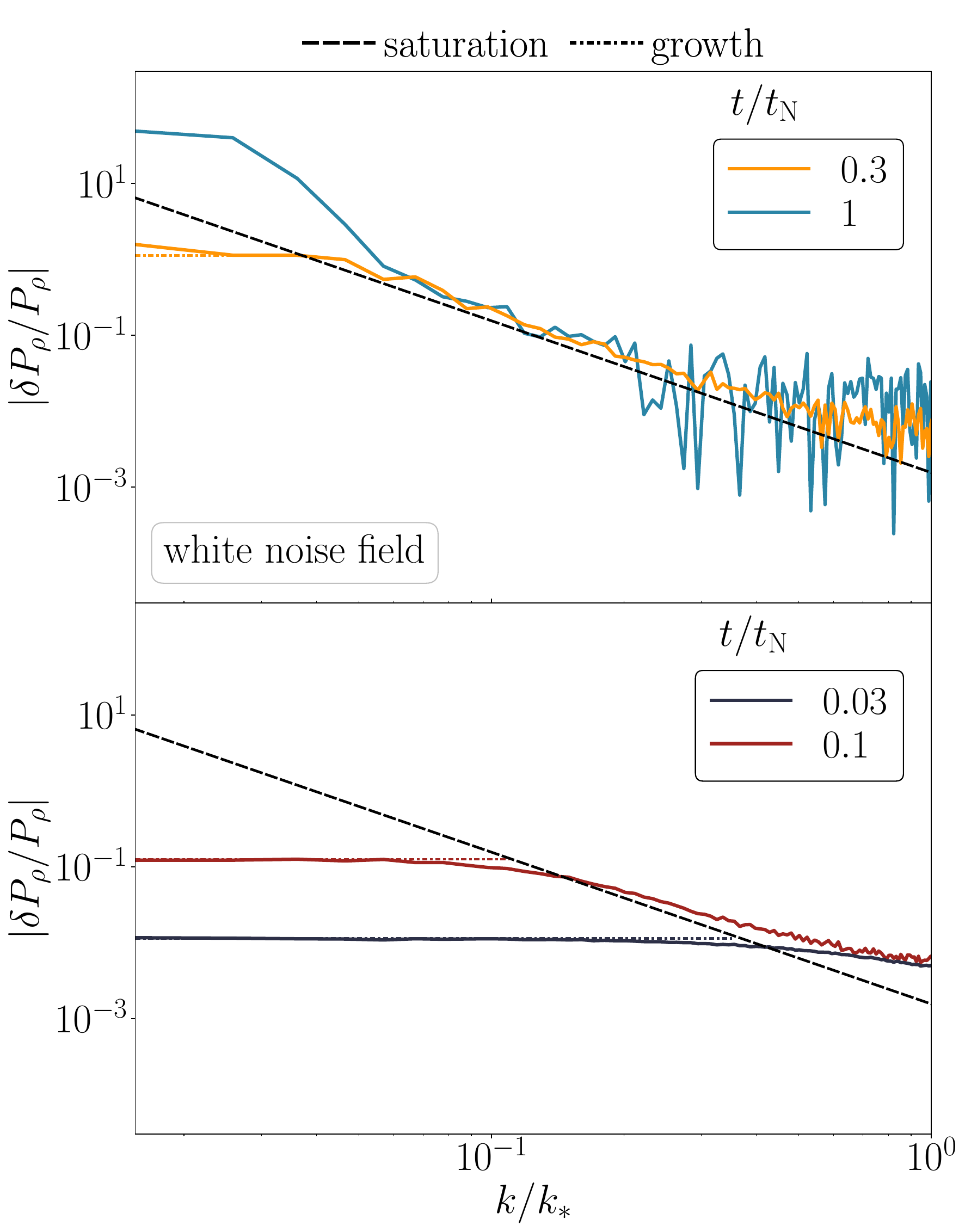}
    \caption{Gravitational growth of the density power spectrum for white noise field simulations (solid lines).  For $k\ll \kp$ the evolution, growth, and saturation is very similar to the shell initial momentum case of Fig.~\ref{fig:dP} for the same parameters $t_\dB/t_N=0.006$ and $\brho/m$.
    }
    \label{fig:dPwhite}
\end{figure}

As a final check on the generality of our results, we consider the case of white noise initial field fluctuations
\begin{equation}
P_i(k) = 
\begin{cases}
{\rm const.} & k\le \kp \\
0 & k>\kp
\end{cases} ,
\end{equation}
with the constant set by the normalization $\psi_{\rm rms}^2=\brho/m$ and with the same $t_\dB/t_\GN$.
For modes where $k\ll \kp$, we would expect the same perturbative scalings to apply whereas modes near $k \sim \kp$ would differ in their behavior due to the effect of field modes on the same scale.   In Fig.~\ref{fig:dPwhite}, we show that indeed the growth and saturation behavior in this regime is very similar to the thin shell case.  \wh{Finally, though we only simulate wave dark matter here, we expect from our dynamical analysis that $N$-body simulations of cold dark matter behave in the same way above the free streaming scale whereas their fluctuations would continue to grow gravitationally on small scales due to the lack of free streaming.}

\section{Discussion}
\label{sec:discussion}

The wave interference of high momentum modes of a hot FDM distribution causes white noise density fluctuations at low wavenumbers. 
We have shown that wave interference effects  prevent the growth of these density fluctuations below the free streaming length.    Above the free streaming scale, these density fluctuations grow on the gravitational dynamical time scale as usual.  Predicted by the wave perturbation theory developed here,  this behavior is verified in \Schro-Poisson simulations with pedagogical examples of a high momentum shell and white noise field configurations.   These results show that hot wave dark matter behaves in the same way as hot particle dark matter in this regard.

Our results provide insight on the gravitational instability of FDM that is either born hot or acquires its large momentum by interactions.  The latter occurs with gravitational interactions themselves in the process of dark matter halo formation for FDM and $m \gtrsim 10^{-22}$\, eV where gravitational infall velocity provides a sufficiently large momentum that the deBroglie wavelength is much smaller than the halo scale.   Here wave interference plays the same role in stabilizing the halo as the coarse grained multistream behavior of particle dark matter in virial equilibrium \cite{Gough:2022pof}. 
\wh{Moreover, we verify that stellar heating bounds from interference are unmodified by gravitational instability
\cite{Dalal:2022rmp}.}

In the self-interaction case, even the small self interaction of the QCD axion can shift its initially cold momentum higher in the early universe and affect the later gravitational stability of small scale structure \cite{Gorghetto:2024vnp}. 

For the evolution of linear density perturbations in an initially hot FDM distribution similar effects occur.  In this case there  is an additional consideration that the dynamical time at the background density is the Hubble time.  For simplicity we have restricted our analysis to  timescales that are short with respect to the Hubble time and also begin the analysis in the matter dominated regime.  The main difference  in both the wave and particle cases is that their momenta redshift and  the free streaming length becomes
\begin{equation}
\lambda_{\rm fs} = \int \frac{d\ln a}{aH } \frac{k}
{\sqrt{k^2 + a^2 m^2}},
\end{equation}
where we have switched to comoving coordinates. 
For  hot particle dark matter, e.g.\ massive neutrinos that decouple while relativistic, the maximal free streaming scale occurs when the particles become non-relativistic and thereafter the continued free streaming per Hubble time decreases in the matter dominated epoch.   Similarly for hot wave dark matter that becomes non-relativistic before matter-radiation equality, the free streaming length logarithmically increases in radiation domination, reaching a maximum total value at equality, and thereafter the subsequent free streaming again decreases in the matter dominated epoch \cite{Hu:2000ke,Amin:2022nlh,Liu:2024pjg}.  
 As with neutrinos, perturbations would continue to grow once the free streaming scale becomes smaller than the wavelength, leading to a gradual suppression below the free streaming scale at matter-radiation equality. This is only smaller than the maximal free streaming scale by a logarithmic factor that depends on the epoch of production \cite{Liu:2024pjg}.
 Thus, instead of a sharp break from growth to saturation exhibited in our non-expanding simulations, we expect a gradual reduction of the initial white noise spectrum to have less high $k$ power beginning at the free streaming scale at equality.
\wh{Thus observational bounds on warm or hot wave dark matter, e.g.\ from the Ly$\alpha$ forest, remain unaffected compared with the cold case \cite{Irsic:2019iff} if the scale is above the maximal free streaming scale, whereas they would be relaxed if below, but in a manner that requires cosmological simulations to quantify.}

Our techniques can readily be extended to \wh{this}
expanding case to test these expectations by perturbing around the adiabatically evolving free wave solution in the WKB approximation.    We leave these considerations to future work.

\bigskip
Note added: upon completion of this project, Ref.~\cite{Amin:2025dtd} appeared which addresses the related question of the gravitational stability of warm particle white noise density fluctuations under its Jeans scale.

\acknowledgments

We thank  Christian  Capanelli, Keisuke Harigaya, Austin Joyce, Evan McDonough for useful conversations. R.L. \& W.H. are supported by U.S.\ Dept.\ of Energy contract DE-FG02-13ER41958 and the Simons Foundation. HX is supported by Fermi Forward Discovery Group, LLC under Contract No.\ 89243024CSC000002 with the U.S.\ Dept.\ of Energy, Office of Science, Office of High Energy Physics.

\vfill
\onecolumngrid
\appendix
\bibliographystyle{apsrev4-2}

\end{document}